\documentclass[a4paper,12pt]{article}

\usepackage{amsmath}
\usepackage{amssymb}
\usepackage{cite}
\usepackage{color}
\usepackage{colordvi}
\usepackage{graphicx}

\makeatletter
 
 \@addtoreset{equation}{section}
\makeatother

\providecommand{\pint}{\makebox[0pt][l]{\hspace{3.4pt}$-$}\int}

\begin{document}

\begin{flushright}
\parbox{4.2cm}
{
 {\tt hep-th/0504209} \hfill \\
 KEK-TH-1001 \\
 OU-HET-524\\
 UTHEP-500  
 }
\end{flushright}

\vspace*{0.4cm}

\begin{center}
 \Large\bf Integrability and Higher Loops \\
in AdS/dCFT Correspondence
\end{center}
\vspace*{0.7cm}
\centerline{\large Yoshiaki Susaki$^{\ast\dagger a}$, 
Yastoshi Takayama$^{\ddagger b}$ 
and Kentaroh Yoshida$^{\ast c}$}
\begin{center}
$^{\ast}$\emph{Theory Division,   	
Institute of Particle and Nuclear Studies, \\
High Energy Accelerator Research 
Organization (KEK),\\
Tsukuba, Ibaraki 305-0801, Japan.} 
\\
$^{\dagger}$\emph{Institute of Physics, University of Tsukuba, \\
Tsukuba, Ibaraki 305-8571, Japan.} 
\\
$^{\ddagger}$\emph{
Department of Particle and Nuclear Physics, \\
The Graduate University for Advanced Studies, \\
Tsukuba, Ibaraki 305-0801, Japan.}
\\
$^{\ddagger}$\emph{
Department of Physics, Osaka University,
Toyonaka, Osaka 560-0043, Japan.} 

\vspace*{0.3cm}
$^a${\tt susaki@post.kek.jp} \quad  
$^c${\tt kyoshida@post.kek.jp} \\
$^b${\tt takayama@het.phys.sci.osaka-u.ac.jp}
\end{center}

\vspace*{0.7cm}

\centerline{\bf Abstract} 

We further study the correspondence between open semiclassical strings and long defect operators which is discussed in our previous work [hep-th/0410139].
We give an interpretation of the spontaneous symmetry breaking of $SO(6)\rightarrow SO(3)_{\rm H} \times SO(3)_{\rm V}$ from the viewpoint of the Riemann surface by following the argument of Minahan. 
Then we use the concrete form of the resolvent for a single cut solution and compute the anomalous dimension of operators dual to  an open pulsating string at three-loop level.
In the string side we obtain the energy of the open pulsating string solution by semiclassical analysis.
Both results agree at two-loop level but we find a three-loop discrepancy.

\vspace*{0.5cm}

\vfill \noindent {\bf Keywords:}~~{\footnotesize AdS/CFT, spin chain,
integrability, semiclassical string, defect CFT}

\thispagestyle{empty}
\setcounter{page}{0}

\newpage 

\section{Introduction and Summary}

The AdS/CFT duality \cite{M} implies the equality between the energies $E$ of quantum string states (as functions of the effective string tension $T= \frac{R^2}{2\pi\alpha'} = \frac{\sqrt{\lambda}}{2\pi}$ and quantum numbers like $S^5$ angular momentum $J^i$) and the dimensions $\Delta$ of the corresponding local SYM operators. 
It is a non-trivial problem to check the relation $E=\Delta$ except 1/2 BPS operators which are dual to the supergravity states.

For generic non-BPS states it has seemed to be difficult to check the duality before the celebrated work of Berenstein-Maldacena-Nastase (BMN) \cite{BMN}, and then Gubser-Klebanov-Polyakov \cite{GKP2} and Frolov-Tseytlin \cite{FT}.  
This type of AdS/CFT duality at non-BPS regimes is investigated in each of subsectors of the full superconformal group $PSU(2,2|4)$\,. 
It is well known that the $SU(2)$ \cite{Beisert:2003ea}, $SL(2)$ \cite{SL2,QCD,Beisert:2003ea,BS,KZ} and $SU(2|3)$ sectors \cite{SU(2|3)} as well as $PSU(2,2|4)$ \cite{BS} are complete closed sectors,
and $SO(6)$ \cite{MZ,BMSZ}, and $SU(3)$ sector \cite{SU(3)} are closed sectors at one-loop.  
In particular, the $SU(2)$ sector is quite well considered and higher-loop contributions are also well studied \cite{Serban:2004jf,Beisert:2004hm}. In this sector the full loop Bethe roots are proposed in \cite{Beisert:2004hm} and the result perturbatively computed in the SYM side is realized from the Bethe roots. 
The SYM result agrees with that in the string side at two-loop level, and the disagreement appears from three loops.

The matching between string and SYM sides is also observed at the action level (coherent state method) as well as at Bethe roots and Bethe equations \cite{Kruczenski}. 
The matching is initially shown by Kruczenski \cite{Kruczenski} in a one-loop analysis of the $SU(2)$ sector. 
The result is extended to the two-loop level in Refs.\ \cite{KRT,ST-coh,KT-coh}.
The matching at the action level is investigated in other sectors such as $SU(3)$ \cite{HL,KT-coh}, $SL(2)$ \cite{ST-coh,Kru-sl2}, $SO(6)$ and sectors \cite{ST-coh,KT-coh}.

The classical integrability of the (super)string on the  AdS$_5$$\times$S$^5$  may also play a key role in this correspondence \cite{BPR,DNW,Arutyunov}.
The matching of the spectra and the integrable structures between the string  sigma models and the spin chains are confirmed up to and including the two-loop level in some cases \cite{AS}.

The correspondence between open string theories and gauge theories are also studied in various setups. 
The correspondence for the near-BPS sectors are discussed in \cite{BGMNN,LP,Imamura:2002wz}.
For the case of the far-from-BPS sectors it is studied in \cite{Open-sol,Chen:2004mu,DM,Susaki:2004tg}.
The open strings on giant gravitons are studied in \cite{giant}.

\bigskip
In our work \cite{Susaki:2004tg} we have considered the defect conformal field theory (dCFT) by considering the defect operators in the setup where the AdS$_4\times$S$^2$ brane (D5-brane) is inserted in the bulk AdS$_5\times$S$^5$ background \cite{DFO}. 
In the dCFT we may consider the defect operators, i.e. the operators with defect fields belonging to the fundamental representation, instead of the trace operation.  
These operators correspond to the states of open string ending on the AdS$_4\times$ S$^2$-brane (D5-brane).  
The matrix of anomalous dimensions for defect operators consisting of $SO(6)$ scalar fields is represented by an integrable open $SO(6)$ spin chain \cite{DM}. 
DeWolfe and Mann obtained the complete defect BMN operators in the scalar sector and showed that those are reduced to the corresponding BMN operators obtained by Lee and Park \cite{LP} when one takes the long operator limit.
In \cite{Susaki:2004tg} we have discussed the coherent state method in this setup. 

\bigskip
In this letter we further study the correspondence of the long defect operators and open string sigma model \cite{Susaki:2004tg} at three-loop level.
We analyze the one-loop anomalous dimension in the $SO(6)$ sector by Bethe ansatz techniques (dCFT side) by following the argument of Minahan \cite{Minahan:2004ds}.
We give an interpretation of the spontaneous symmetry breaking, $SO(6)\rightarrow SO(3)_{\rm H}\times SO(3)_{\rm V}$ from the viewpoint of the Riemann-Hilbert problem in the thermodynamic limit.
We concretely construct the resolvent for a single cut solution corresponding to an open pulsating string solution.
We confirm that both Bethe ansatz techniques and coherent state method \cite{Susaki:2004tg} give the same one-loop anomalous dimension of the operator dual to the open pulsating string.
In addition, two-loop and three-loop contributions to the resolvent are evaluated by using the knowledge for the Inozemtsev spin chain \cite{Serban:2004jf}, and the anomalous dimension is calculated at three-loop level.
This is consistent with the closed string result, which is seen through the doubling trick.

In the string side we obtain the energy of the open pulsating string including higher-loop corrections by following the solution ansatz \cite{Susaki:2004tg}.
We see the agreement between the dCFT and the open string side at two loops and a three-loop discrepancy.
In addition we compute the period of the open pulsating string.
We see that the doubling trick formula holds for the period too.
\bigskip

This letter is organized as follows:
In section 2, the one-loop anomalous dimension of the operator dual to the open pulsating string is analyzed.
We give an interpretation of the spontaneous symmetry breaking, $SO(6)\rightarrow SO(3)_{\rm H}\times SO(3)_{\rm V}$.
We see that the anomalous dimension computed from the resolvent at one-loop level agrees with our previous result \cite{Susaki:2004tg}.
By extending the analysis the anomalous dimension at three-loop level is obtained.
In section 3, we compute the energy of open pulsating string and the frequency of the string at the order of $(\lambda/L^2)^6$.
We find the three-loop discrepancy between dCFT and string side.

\bigskip
After we completed the manuscript we received an interesting work \cite{McLoughlin:2005gj}, where the issues  related to our work are discussed.

\section{Anomalous Dimension of Long Defect Operators}

\subsection{One-Loop Analysis and Symmetry Breaking}
In the dCFT \cite{DFO} the dual defect operators for open strings are \cite{DM,Susaki:2004tg}
\begin{eqnarray}
\mathcal{O} = \psi_{m,j_1,\ldots,j_L,n}\,\bar{q}_mX^{j_1}\cdots
 X^{j_L}q^n\,. 
\label{cop}
\end{eqnarray}
The one-loop anomalous dimension matrix $\displaystyle \Gamma_{\cal O}$ for (\ref{cop}) is obtained in the work of \cite{DM}
\begin{eqnarray}
\Gamma_{\cal O} &=& \Gamma^{\rm bulk}_{\mathcal{O}} + 
\Gamma^{\rm defect}_{\mathcal{O}} \nonumber \\
&=&  \frac{\lambda}{16\pi^2}
\sum_{l=1}^{L-1}H_{l,l+1} + \frac{\lambda}{16\pi^2}\Bigl[
(2I_{\bar{q}1} +2\bar{S}_{\bar{q}1}) 
+ (2I_{Lq} +2 S_{Lq})
\Bigr]\,, 
\label{open-spin} \\
H_{l,l+1} &=& K_{l,l+1} + 2I_{l,l+1} -2 P_{l,l+1}\,. 
\end{eqnarray}
The bulk part (the first term in (\ref{open-spin})) is the same 
as the result of Minahan and Zarembo \cite{MZ}.
The defect part (the second term in (\ref{open-spin})) is written in terms of
\begin{eqnarray}
\begin{array}{@{\,}cl@{\,}}
I^{\bar{m} I}_{\bar{n} J} = \delta^{\bar{m}}_{\bar{n}} 
\delta_{J}^{I}\,, \quad 
I^{I m}_{J n} = \delta_{J}^{I}\delta^{m}_{n}\,, \quad
S_{Jn}^{I m} = -i\epsilon_{IJK}\sigma^K_{nm}\,, \quad 
\bar{S}_{\bar{n}J}^{\bar{m}I} = i\epsilon_{IJK}\sigma^K_{\bar{m}\bar{n}}\,,
& \big(\mbox{for $SO(3)_{\rm H}$}\big) \\
I_{\bar{n}B}^{\bar{m}A}=\bar{S}_{\bar{n}B}^{\bar{m}A} = \delta_{\bar{n}}^{\bar{m}}\delta^A_B\,,
\quad
I_{Bn}^{Am} =S_{Bn}^{Am} = \delta^m_n\delta^A_B\,,
& \big(\mbox{for $SO(3)_{\rm V}$}\big)
\end{array}
\end{eqnarray}
and the others vanish.
Here we should note that the bulk part has no periodicity, but the integrable boundary arising from the defect contribution ensures the integrability of the spin chain.
\bigskip
\par
From now on we will discuss the spontaneous symmetry breaking 
$SO(6) \rightarrow SO(3)_{\rm H} \times SO(3)_{\rm V}$ due to the
presence of an AdS$_4$$\times$S$^2$-brane from the viewpoint of the 
Bethe equation and Riemann surface.  
By the standard argument the diagonalization problem of the anomalous dimension matrix (\ref{cop}) reduces to solving problem of the $SO(6)$ open Bethe equations 
\begin{eqnarray}
&& \hspace*{-1cm} 
\left(\frac{u_{1i}+i/2}{u_{1i}-i/2}\right)^{2L} = 
 \prod_{j (\neq i)}^{n_1} \frac{u_{1i}-u_{1j}+i}{u_{1i}-u_{1j}-i}
 \prod_{j}^{n_2} \frac{u_{1i}-u_{2j}-i/2}{u_{1i}-u_{2j}+i/2}
 \prod_{j}^{n_3} \frac{u_{1i}-u_{3j}-i/2}{u_{1i}-u_{3j}+i/2} \nonumber
\\
&& \hspace*{2.5cm} \times
 \prod_{j (\neq i')}^{n_1} \frac{u_{1i}+u_{1j}+i}{u_{1i}+u_{1j}-i}
 \prod_{j}^{n_2} \frac{u_{1i}+u_{2j}-i/2}{u_{1i}+u_{2j}+i/2}
 \prod_{j}^{n_3} \frac{u_{1i}+u_{3j}-i/2}{u_{1i}+u_{3j}+i/2}\,,
\label{Bethe1-1}
\\
&& 1 = 
 \prod_{j (\neq i)}^{n_2} \frac{u_{2i}-u_{2j}+i}{u_{2i}-u_{2j}-i}
 \prod_{j} \frac{u_{2i}-u_{1j}-i/2}{u_{2i}-u_{1j}+i/2} 
 \prod_{j (\neq i')}^{n_2} \frac{u_{2i}+u_{2j}+i}{u_{2i}+u_{2j}-i}
 \prod_{j} \frac{u_{2i}+u_{1j}-i/2}{u_{2i}+u_{1j}+i/2}\,,\label{Bethe1-2}
\\
&& 1 =
 \prod_{j (\neq i)}^{n_3} \frac{u_{3i}-u_{3j}+i}{u_{3i}-u_{3j}-i}
 \prod_{j} \frac{u_{3i}-u_{1j}-i/2}{u_{3i}-u_{1j}+i/2} 
 \prod_{j (\neq i')}^{n_3} \frac{u_{3i}+u_{3j}+i}{u_{3i}+u_{3j}-i}
 \prod_{j} \frac{u_{3i}+u_{1j}-i/2}{u_{3i}+u_{1j}+i/2}\,,
\label{Bethe1-3}
\end{eqnarray} 
where $\prod_{j (\neq i')}$ is a product over $j$ except $u_i + u_j =0$\,.
Note that the above Bethe Eq. (\ref{Bethe1-1}) describes an open string as a closed string with only a right-moving or left-moving mode through the doubling trick, and the energy of the open spin chain obtained from the Bethe Eq. (\ref{Bethe1-1}) is twice of the anomalous dimension for an open string in the usual sense. This kind of treatment is needed to use the periodicity as in the case of closed strings. 

\par
In general, in the case of open strings, one has to take into account of boundaries by including factors denoting reflections. 
However, we are interested in the open string case where we may rewrite an open string as a closed string with only one mode, not two by assuming the doubling trick.
Namely, the analysis should be reduced to a periodic case.
So we will set the reflection factors to be $1$.
In fact, in the analysis for BMN operators \cite{DM} the reflection factors become 1 and the presence of boundaries does not affect the anomalous dimensions.

The anomalous dimension, which is half of the open spin chain energy $E$, is given by
\begin{eqnarray}
\gamma
 =
\frac{E}{2}
 =
\frac{\lambda}{4\pi^2}
 \sum_{j=1}^{n_1} \frac{1}{u^2_{1j}+\frac{1}{4}}.
\label{anomalous01}
\end{eqnarray}

Following the argument in \cite{Minahan:2004ds} we impose the following four ansatz:
\begin{eqnarray*}
 &\mbox{$\bullet$}& \mbox{half filling condition $n_2=n_3=n_1/2$.} \label{half-fill}\\
 &\mbox{$\bullet$}& \mbox{the Bethe root distributions of $u_2$ and $u_3$ are the same.}\\
 &\mbox{$\bullet$}& \mbox{the $u_{1}$ roots are on multiple cuts $C_l$, while the $u_{2}$ and $u_{3}$ are on a single cut $C'$}.\\
 &\mbox{$\bullet$}&
\mbox{A root with flipped sign has also to be a root.}
\end{eqnarray*}
Let us rescaling $u=2Lx$ and take the thermodynamic limit, $L \to \infty$.
In this limit the Bethe equations (\ref{Bethe1-1})-(\ref{Bethe1-3}) and the anomalous dimension (\ref{anomalous01})  are reduced to integral forms
\begin{eqnarray}
 \frac{1}{x} + 2n_l\pi
&=&
 2 \pint_{C_l} dx'
 \frac{\sigma(x')}{x-x'}
 +2\sum_{k(\neq l)} 
 \int_{C_k} dx' \frac{\sigma(x')}{x-x'}
 -2\int_{C'}dx' \frac{\rho(x')}{x-x'},
\quad
  x \in C_l
 \label{IntEq01}
\\
 0
&=&
 2\pint_{C'}dx' 
 \frac{\rho(x')}{x-x'}
-\sum_k \int_{C_k}dx'  \frac{\sigma(x')}{x-x'},
\quad
 x \in C'\label{IntEq02}
\end{eqnarray}
and
\begin{equation}
\gamma
=
 \frac{\lambda}{16\pi^2 L} 
 \sum_{k}
\int_{C_k} \frac{\sigma(x)}{x^2} dx
 \label{ADInt}
\end{equation}
where $n_l$ labels the log branch.
Here we have introduced the Bethe root densities 
\begin{eqnarray}
 \sigma(x)
\equiv
 \frac{1}{L} \sum_{j=1}^{n_1} \delta(x-x_{1j}),
\quad 
 \rho(x)
\equiv
 \frac{1}{L} \sum_{j=1}^{n_2} \delta(x-x_{2j})
=
 \frac{1}{L} \sum_{j=1}^{n_3} \delta(x-x_{3j}).
\end{eqnarray}
The root densities are constrained by a normalization condition:
\begin{eqnarray}
 2 \int_{C'} \rho(x) dx
=
 \sum_{k} \int_{C_k}  \sigma(x) dx
=
\frac{n_1}{L},
 \label{normal}
\end{eqnarray}
where we have used the half-filling condition.

\par
By using Hilbert transformation with (\ref{IntEq02}) we  obtain
\begin{eqnarray}
\lefteqn{\rho(x)
=
-\frac{1}{2\pi^2} \sqrt{(x-a)(x-\bar{a})}
 \pint_a^{\bar{a}} \frac{dx'}{x-x'}
 \frac{1}{\sqrt{(x'-a)(x'-\bar{a})}}
 \sum_k \int_{C_k}dx''
 \frac{\sigma(x'')}{x'-x''}} \label{rho1}
\hspace*{15cm}
\\
\lefteqn{=
 -\frac{1}{2\pi i} \sum_{k}
 \int_{C_k} dx' \frac{\sigma(x')}{x-x'}
 \sqrt{\frac{(x-a)(x-\bar{a})}{(x'-a)(x'-\bar{a})}},}
 \label{rho2}
\hspace*{14.1cm}
\end{eqnarray} 
where we set the endpoints of $C'$ as $a$ and its complex  conjugate $\bar{a}$.

\par

We substitute (\ref{rho1}) into (\ref{normal}), then
\begin{eqnarray}
 \int_{C'} \rho(x) dx
=
  \frac{1}{2} \sum_k \int_{C_k}  \sigma(x) dx
 -\frac{1}{4} \sum_k \int_{C_k} dx
  \sigma(x) \frac{a+\bar{a}-2x}{\sqrt{(x-a)(x-\bar{a})}}. 
   \label{intrho}
\end{eqnarray}
We see that the second term in the right hand side of (\ref{intrho}) must vanish because of the consistency between (\ref{intrho}) and (\ref{normal}).
This can be achieved by taking a limit
\begin{eqnarray}
\mbox{Im}\,a \to \infty \quad \mbox{and} \quad \frac{\mbox{Re}\,a}{\mbox{Im}\,a} \to 0. \label{lim}
\end{eqnarray}
Since $a$ and $\bar{a}$ are the endpoints of $C'$, in the limit of (\ref{lim}) the cut $C'$ lies on the imaginary axis of $x$ ($C'\to i\mathbb{R}$).
Hence the contour $C'$ splits the complex plane into two regions.
In addition $\sqrt{\frac{(x-a)(x-\bar{a})}{(x'-a)(x'-\bar{a})}}$ becomes a sign function $\mbox{sign}(x,x')$,  which gives the $+$($-$) sign if $x$ and $x'$ are on the same (opposite) sides of $C'$.
From (\ref{rho2}) we obtain, in this limit, 
\begin{eqnarray}
 \int_{C'} dx' \frac{\rho(x')}{x-x'} =  \sum_{k'} \int_{C_{k'}} dx' \frac{\sigma(x')}{x-x'}, \label{opposite}
\end{eqnarray}
where the sum over the index $k'$ refers to cuts on the opposite side of $C'$ from $x$.

By taking (\ref{opposite}) into account, (\ref{IntEq01}) is separated into two equations:
\begin{eqnarray}
 \frac{1}{x} + 2 \pi n_{l_+}
&=&
 2 \pint_{C_{l_+}} \frac{\sigma(x')}{x-x'} dx'
 +2 \sum_{k_+(\neq l_+)} \int_{C_{k_+}} 
 \frac{\sigma(x')}{x-x'} dx', 
 \quad x \in C_{l_+}
 \label{IntEq+} 
\\
 \frac{1}{x} + 2 \pi n_{l_-}
&=&
 2 \pint_{C_{l_-}} \frac{\sigma(x')}{x-x'} dx'
 +2 \sum_{k_-(\neq l_-)} \int_{C_{k_-}} 
 \frac{\sigma(x')}{x-x'} dx', 
 \quad x \in C_{l_-}
 \label{IntEq-} 
\end{eqnarray}
with the $+(-)$ specifying the Bethe roots on the right(left) hand side of $C'$. 
Each side of the Bethe root distributions can be determined independently, and hence this system has two independent sets of roots.
It is worthwhile noting that the absence of the cyclicity of trace in the defect operators ${\cal O}$ implies that two Bethe roots are {\it completely independent}, while for single trace operators they are indirectly correlated \cite{Minahan:2004ds}.

\bigskip

We give a physical interpretation of the limit (\ref{lim}) in terms of D-brane.
In our setup the insertion of a D5-brane into a stack of D3-branes breaks the $SO(6)$ R-symmetry down to  $SO(3)_{\rm H} \times SO(3)_{\rm V}$.
Each of $SO(3)_{\rm H}$ and $SO(3)_{\rm V}$ should correspond to each side of Bethe roots and hence we identify the breaking of the R-symmetry with that of the complex plane.
In this interpretation the {\it diagonal} R-symmetry breaking allows us to treat the two Bethe roots {\it independently}.
We conclude that the limit (\ref{lim}) corresponds to the D5-brane insertion into the D3-branes.

\begin{figure}[Htbp]
\begin{center}
 \begin{minipage}{0.45\hsize}
  \begin{center}
   \includegraphics[width=70mm]{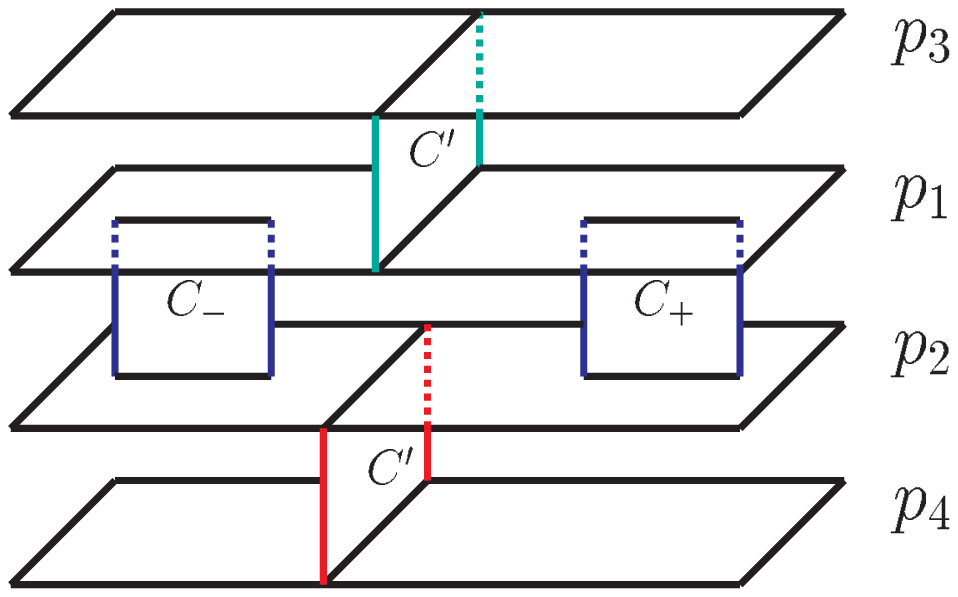}
  \end{center}
  \caption{The branch cuts for the open pulsating solution}
  \label{fig:one}
 \end{minipage}
\hspace*{1cm}
 \begin{minipage}{0.45\hsize}
  \begin{center}
   \includegraphics[width=70mm]{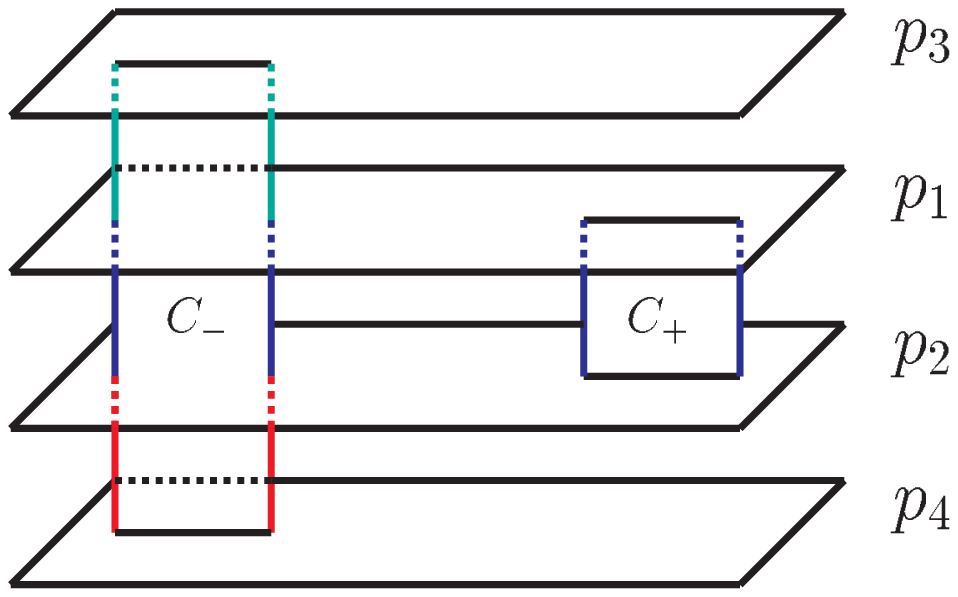}
  \end{center}
  \caption{The removal of the cuts $C'$. The cut $C_-$ connects $p_3$ and $p_4$, while the cut $C_+$ does $p_1$ and $p_2$.}
  \label{fig:two}
 \end{minipage}
\end{center}
\end{figure}

We also argue the limit (\ref{lim}) from the viewpoint of algebraic curves \cite{Beisert:2004ag,KZ,alg,Kazakov:2004qf} by introducing $p_i$ ($i=1\cdots 4$):
\begin{eqnarray}
 p_1(x) &\equiv& G_1(x) - G_3(x) +\frac{1}{2x}, \qquad p_2(x) \equiv G_2(x) - G_3(x) -\frac{1}{2x}, \\
 p_3(x) &\equiv& G_3(x) + \frac{1}{2x}, \qquad p_4(x) \equiv -G_2(x) -\frac{1}{2x},
\end{eqnarray}
with
\begin{eqnarray}
 G_1(x) &\equiv& \sum_k \int_{C_k} dx' \frac{\sigma(x)}{{x-x'}}, \qquad
 G_2(x) = G_3(x) \equiv \int_{C'} dx' \frac{\rho(x)}{x-x'}.
\end{eqnarray}
Then the SO(6) open Bethe equations become
\begin{eqnarray}
 \begin{array}{ll}
 \displaystyle  \not{p}_3(x)-\not{p_1}(x) = 0 &  x\in C', \\
 \displaystyle  \not{p}_1(x)-\not{p_2}(x) = 2\pi n_{l\pm} &  x \in C_{l_\pm}, \\
 \displaystyle  \not{p}_2(x)-\not{p_4}(x) = 0 & x \in C',
 \end{array}
\end{eqnarray}
where $\not{p}_i$ are the principal values on each of cuts: $\not{p}_i$ $=\frac{1}{2}p_i(x-\epsilon)+\frac{1}{2}p_i(x+\epsilon)$.
As the cut $C'$ lies on the imaginary axis of $x$, this situation is similar to the case of closed pulsating string \cite{Engquist:2003rn} \cite{Beisert:2004ag}, where the branch cuts of $C'$s and $C_\pm$ are drawn in Fig. \ref{fig:one}.
By removing the branch cuts $C'$ we obtain the Fig.\ref{fig:two}, where the two cuts $C_+$ and $C_-$ are  completely independent.
This is consistent with the two completely independent equations (\ref{IntEq+}) and (\ref{IntEq-}).

\bigskip

It is useful to introduce resolvents for solving the integral equations.
When we consider the $SO(3)_{\rm H}$ sector only, the resolvents should be set to
\begin{eqnarray}
 G_+(x)
=
 \sum_{k_+} \int_{C_{k_+}}  dx' \frac{\sigma(x')}{x-x'} ,
\qquad
G_-(x)=0.
\end{eqnarray}
Then the integral equation (\ref{IntEq+}) becomes
\begin{eqnarray}
&& G_+(x+i0) + G_+(x-i0) = \frac{1}{x} + 2 \pi n_{l_+},
\qquad
 x \in C_{l_{+}}.
\end{eqnarray}
\par
Let us consider the simplest solution, that is, a single cut solution.
We set by 
\begin{eqnarray}
 G_+(0)= \pi n_+,
\end{eqnarray}
%
then the resolvent is
\begin{eqnarray}
 G_+(x)
=
 \frac{1}{2x}
 -\frac{1}{2x}\sqrt{(2\pi n_+ x)^2 +1 } + \pi n_+.
\end{eqnarray}
%
%
We obtain the anomalous dimension from (\ref{ADInt})
\begin{eqnarray}
 \gamma
=
 -\frac{\lambda}{16 \pi^2 L}G'_+(0)
=
 \frac{\lambda}{16L} n_+^2.
\end{eqnarray}
This result {\rm completely} agrees with  the previous one in \cite{Susaki:2004tg}.

\subsection{Higher-Loop Analysis}
Let us discuss the defect part of the anomalous dimension.
By symmetry argument the defect part of the anomalous dimension including higher-loop corrections has a similar  flavor structure to the one-loop one:
\begin{eqnarray}
\Gamma_{\cal O}^{\rm defect-left}
&=&
\bigg( \frac{\lambda}{8\pi^2} + C_{12} \lambda^2 +  C_{13} \lambda^3  +\cdots \bigg)
\delta_{J_2}^{I_2}
\left[\delta^m_n \delta^{J_1}_{I_1} - i \epsilon_{{I_1}{J_1}K}\sigma^K_{mn} \right] \nonumber\\
&+&\,\bigg(C_{22} \lambda^2 + C_{23} \lambda^3  +\cdots \bigg) \delta^{I_1}_{J_1} \left[\delta^m_n \delta_{I_2}^{J_2} - i \epsilon_{{I_2}{J_2}K}\sigma^K_{mn} \right] \nonumber\\
&+&\cdots,
\end{eqnarray}
and similarly for $\Gamma_{\cal O}^{\rm defect-right}$.
(Here ``$\cdots$'' denotes the $SO(3)_{\rm V}$ contribution and the contribution from fermions, which are irrelevant to our consideration after taking the large $L$ limit.)
Hence we may expect that the boundary contributions would be absorbed into the bulk one in the coherent state method \cite{Susaki:2004tg}.
This expectation implies that the reflection factors in the higher-loop open Bethe equations would be also set $1$ when we consider only in the $SO(3)_{\rm H}$ sector. 
%

It is straightforward to extend the analysis in the previous section to three-loop level by the above argument and  following the argument in \cite{Serban:2004jf,Minahan:2004ds}.
Then the three-loop anomalous dimension is
\begin{eqnarray}
 \gamma=2LT \int dx \left(\frac{1}{x^2} +\frac{3T}{x^4} +\frac{10T^2}{x^6}\right) \sigma(x),
\label{intADM3}
\end{eqnarray}
and the integral equation is
\begin{eqnarray}
 \frac{1}{x} + \frac{2T}{x^3} + \frac{6T^2}{x^5} +2\pi n_{l_+}
=
 2 \pint_{C_{l_+}} \frac{\sigma(x')}{x-x'} dx'
 +2 \sum_{k_+(\neq l_+)} \int_{C_{k_+}} 
 \frac{\sigma(x')}{x-x'} dx', 
 \quad x \in C_{l_+}.
\end{eqnarray}
Here we have defined as $\displaystyle T \equiv \frac{\lambda}{64 \pi^2 L^2}$.
Then we obtain the anomalous dimension at the three-loop level
\begin{eqnarray}
 \gamma 
&=&
 \frac{n^2 \lambda}{16L}
 -\frac{n^4 \lambda^2}{1024 L^3}
 +O(\lambda^4).
\label{anomalous-3}
\end{eqnarray}
This is consistent with the closed spin chain result \cite{Minahan:2004ds} through the doubling trick formula ($\kappa\equiv(1-\alpha_{\rm c})^2$)
\begin{eqnarray*}
 \gamma
&=&
 \frac{1}{2}\gamma_{\rm closed}(L_{\rm c})
=
 \frac{1}{2} \bigg[
 \frac{\lambda n_{\rm c}^2 (1-\kappa)}{4L_{\rm c}} - \frac{\lambda^2 n_{\rm c}^4 (1-\kappa)(1+3\kappa)}{64 L_{\rm c}^3}
+\frac{\lambda^3 n_{\rm c}^6 (1-\kappa)(1+3\kappa)\kappa}{128L_{\rm c}^5} \bigg]
\end{eqnarray*}
with $L_{\rm c}=2L$, $\alpha_{\rm c}=1$ and $n_{\rm c}=n$.
Hence the three-loop scaling dimension $\Delta$ of the operator dual to open pulsating string is
\begin{eqnarray}
\frac{\Delta}{L}= \frac{1}{L}\left[ L+ \gamma \right]
=
 1+
 \frac{n^2 \lambda}{16L^2}
 -\frac{n^4 \lambda^2}{1024 L^4}
 +O(\lambda^4). \label{scaling}
 \end{eqnarray}
\par
We note that the three-loop corrections vanish.
In the next section we will evaluate the energy of the open pulsating string.

\section{Semiclassical Energy for Open Pulsating String}

We shall compute the energy of an open pulsating string solution by using the semiclassical method employed in \cite{KT-coh}. 
We consider the action of an open string on $R$$\times$S$^2$ in terms of the spherical coordinates with AdS$_5$ time $t$\,,
\begin{eqnarray}
I &=& -\frac{\sqrt{\lambda}}{4}\int d \tau\! \int_0^\pi \frac{d \sigma}{\pi}\,
\left[\kappa^2 + \partial_a \theta
 \partial^a \theta + \sin^2 \theta \partial_a \phi \partial^a \phi \right]\,.
\end{eqnarray} 
Here we have utilized the conformal gauge $g_{ab}={\rm diag}(-1,1)$ with
$t=\kappa \tau $\,. The solution ansatz we are interested in is 
\begin{eqnarray}
\label{sa}
 \theta = \theta(\tau)\,, \qquad \phi = m \sigma \quad (m\in\mathbf{Z})\,,
\end{eqnarray} 
and it corresponds to the pulsating solution as we will see later
\footnote{Strictly speaking, the boundary conditions are not satisfied, but those are satisfied in the $L\to \infty$ limit.}.
Under this solution ansatz (\ref{sa}), the only non-trivial equation of motion is
\begin{eqnarray}
 \partial_\tau^2 \theta + m^2 \sin\theta \cos\theta = 0\,. 
\label{eom} 
\end{eqnarray} 
The conformal gauge constraint is written as  
\begin{eqnarray}
\label{cgc}
 (\partial_\tau \theta)^2 + m^2 \sin^2\theta = \kappa^2\,, 
\end{eqnarray} 
but it is equivalent to the e.o.m.~(\ref{eom}) since the constraint
(\ref{cgc}) is re-derived by integrating (\ref{eom}) and taking
$\kappa^2$ as an integration constant.

Let us now introduce the action variable $I_{\theta}$ with respect to $\theta$\,. 
The canonical momentum of $\theta$ is $P_\theta = (\sqrt{\lambda}/(2\pi))
\partial_\tau \theta$\,, and $I_{\theta}$ is defined as a contour
integral as follows: 
\begin{eqnarray}
I_\theta \equiv \oint d\theta\, P_\theta
= 2\sqrt{\lambda}\int_{0}^{\pi}\!\!\! d\theta\, 
\sqrt{1 - \frac{m^2}{\kappa^2} \sin^2\theta}\,.
\end{eqnarray} 
When $ L \equiv I_\theta/2 $ and $E\equiv \sqrt{\lambda}\,\kappa/2$ are introduced, the energy density is given by
\begin{eqnarray}
\label{2.6}
\frac{L}{E} = \frac{1}{\pi}\int_{0}^{\pi}\!\!\!d\theta\, 
\sqrt{1 - \frac{m^2 \lambda}{4 E^2} \sin^2 \theta}\,  
= \frac{2}{\pi}\mathbf{E}\left(\frac{m^2\lambda}{4E^2}\right)\,,
\end{eqnarray} 
where we have introduced the complete elliptic integral of the second kind  
\[
\mathbf{E}(k) = \int^{\pi/2}_0\!\!\!\!\!d\varphi\, \sqrt{1-k\sin^2\varphi}\,.
\]
The equation (\ref{2.6}) can be iteratively solved for $E$ by using
the Mathematica, and the energy density $E/L$ is evaluated as  
\begin{eqnarray}
 \frac{E}{L} &=&
 1 + \frac{\lambda m^2}{16 L^2} - \frac{\lambda^2 m^4}{1024 L^4} 
+ \frac{\lambda^3 m^6}{16384 L^6} \nonumber\\
&&
- \frac{13 \lambda^4 m^8}{4194304 L^8}
 + \frac{17 \lambda^5 m^{10}}{67108864 L^{10}} 
- \frac{71 \lambda^{6}m^{12}}{4294967296L^{12}} + O(\lambda^7). \label{energy}
\end{eqnarray}
The first order term in terms of $\lambda$ (or $g$) completely agrees with the energy of open pulsating string energy computed in terms of the Landau-Lifshitz type sigma model \cite{Susaki:2004tg}. 
By replacing $L$ by $L/2$ the closed pulsating string energy is recovered \cite{KT-coh}.
The results in both open and closed cases are related through the doubling trick \cite{Open-sol}.
We find a three-loop discrepancy when we compare (\ref{scaling}) with (\ref{energy}).

\paragraph*{The period of the open pulsating string solution}
Here we comment on the motion of the open pulsating string solution.
The solution of (\ref{eom}) is
\begin{eqnarray}
 \sin \theta(t) = - \mathbf{sn} \left(\frac{\kappa}{m}t\right)
\qquad (t= \kappa \tau)\,, \label{sol}
\end{eqnarray}
where $\mathbf{sn}\left(\frac{\kappa}{m}t\right)$ is one of Jacobi's elliptic function with the period: $ T = 4 \mathbf{K}\left(\frac{m^2}{\kappa^2}\right) $.
Then the frequency of the motion of Eq. (\ref{sol}) is
\begin{eqnarray*}
\omega &=& \frac{2 \pi}{T} =
 \frac{\pi}{2 \mathbf{K}\left(\frac{ \lambda m^2}{4 E^2(\lambda, L)}\right) }
\qquad
\left(\mathbf{K}\left(\frac{m^2}{\kappa^2}\right) = \int_{0}^{\pi/2}\!\!\!\!d\theta\, 
\frac{1}{ \sqrt{1- \frac{m^2}{\kappa^2} \sin^2\theta \,}} \right)
\\
&=&
 1 - \frac{m^2\,\lambda }{16\,L^2} + \frac{3\,m^4\,{\lambda }^2}{1024\,L^4} - 
  \frac{5\,m^6\,{\lambda }^3}{16384\,L^6} \\
&&+ \frac{91\,m^8\,{\lambda }^4}{4194304\,L^8} - 
  \frac{153\,m^{10}\,{\lambda }^5}{67108864\,L^{10}}
+ \frac{781\,m^{12}\,{\lambda }^6}{4294967296\,L^{12}} + {O}(\lambda ^7) \\
\end{eqnarray*} 
Note that the doubling trick formula holds for the period of the motion
by replacing $L$ by $L/2$ \cite{KT-coh}.
An alternative definition of the frequency,
$ \omega = 2 \frac{\partial E}{\partial I_\theta} 
= \frac{\partial}{\partial L}  E(\lambda, L)\,,  
$
also gives the same result.

\section*{Acknowledgments}
We would like to thank H.~Fuji and A.~Yamaguchi for useful discussion.
The work of Y.~T.\ is supported in part by the 21st century COE Program ``Towards a New Basic Science ; Depth and Synthesis'' from the Ministry 
of Education, Culture, Sports, Science and Technology (MEXT) of Japan.
The work of K.~Y.\ is supported in part by JSPS Research Fellowships for Young Scientists.

\end{document}